# Stuck-at Faults in ReRAM Neuromorphic Circuit Array and their Correction through Machine Learning

Vedant Sawal
*M-PAC Lab*
*San Jose State University*
San Jose, USA
vedant.sawal@sjsu.edu

Hiu Yung Wong*
*M-PAC Lab*
*San Jose State University*
San Jose, USA
hiuyung.wong@sjsu.edu

*Abstract—* In this paper, we study the inference accuracy of the Resistive Random Access Memory (ReRAM) neuromorphic circuit due to stuck-at faults (stuck-on, stuck-off, and stuck at a certain resistive value). A simulation framework using Python is used to perform supervised machine learning (neural network with 3 hidden layers, 1 input layer, and 1 output layer) of handwritten digits and construct a corresponding fully analog neuromorphic circuit (4 synaptic arrays) simulated by Spectre. A generic 45nm Process Development Kit (PDK) was used. We study the difference in the inference accuracy degradation due to stuck-on and stuck-off defects. Various defect patterns are studied including circular, ring, row, column, and circular-complement defects. It is found that stuck-on and stuck-off defects have a similar effect on inference accuracy. However, it is also found that if there is a spatial defect variation across the columns, the inference accuracy may be degraded significantly. *We also propose a machine learning (ML) strategy to recover the inference accuracy degradation due to stuck-at faults. The inference accuracy is improved from 48% to 85% in a defective neuromorphic circuit.*

## I. Introduction

Artificial intelligence (AI) is revolutionizing almost every aspect of our society [1]. Its advancement from machine learning to deep learning to foundation models makes it more and more versatile [1]. As it becomes more powerful, it is also more power-hungry. For example, ChatGPT-3 consumes about 10 gigawatt-hour (GWh) of energy to be trained and 1GWh daily to handle hundreds of millions of queries [2]. Edge computing and Internet-of-Things (IoT) using AI are also becoming very popular. Computing at edge devices (such as cell phones and smart watches) and IoT obviates data transmission and protects privacy. Edge devices have very stringent requirements on power consumption and device reliability.

Regardless of the new algorithms and advancements, most AI is based on neural networks (NN) in which Vector-Matrix-Multiplication (VMM) is involved. VMM is energy-consuming and time-consuming due to the movement of data between the memory and the computational units in addition to its computation complexity [3][4]. Therefore, reducing the energy consumption and increasing the speed of VMM (or improving the energy-speed trade-off) is very critical.

Compute-in-Memory (CiM), which stores the data and performs the computation in memory, is promising in solving the aforementioned problem [5]-[7]. VMM is performed through analog computing by applying a voltage across a memory, which stores the synaptic weight of the neural network as its resistance. Through Kirchhoff's voltage and current laws, the resulting current naturally represents the result of VMM. This obviates the need for frequent data transfer and complex digital hardware. One VMM can be finished in one single time step. Emerging memories such as Resistive Random-Access-Memory (ReRAM) [8][9], Spin Transfer Torque Random-Access-Memory (STT-RAM) [10], Phase Change Memory (PCM) [11], etc. are the most promising memory elements for CiM because they are non-volatile and suitable for IoT and edge computing applications.

In this paper, a fully analog ReRAM neuromorphic circuit pre-loaded with training weights is studied. This is expected to be important in IoT and edge devices for inference (such as image recognition). A fully analog circuit is also expected to be more energy and area-efficient because it eliminates the digital circuits (such as adders, registers, shifters, multiplexers, and analog-to-digital converters) between the VMM arrays in which the synaptic weights are stored [12]. Therefore, this paper studies fully analog neuromorphic circuits.

The reliability of fully analog ReRAM neuromorphic circuits is of paramount importance in IoT. Over the lifetime, the ReRAM resistance will change [13], which will degrade its inference accuracy and, thus, its usable lifetime. Moreover, IoT used in harsh environments might experience damage in the ReRAM or the metal wires and contacts (e.g. radiation damage [14][16]). It is important to understand how the spatial distribution of the damage will affect its lifetime. Moreover, it is inevitable to have stuck-at faults in ReRAM after fabrications [17][18]. The ReRAM might be stuck at a low resistance state (stuck-on), a high resistance state (stuck-off), or at a certain resistive value. This reduces the yield and increases the cost, which is not desirable for IoT applications.

The degradation of inference accuracy due to stuck-off fault and resistance drifting in the ReRAM neuromorphic circuit has been investigated [19]. In this paper, stuck-on faults are studied and compared to stuck-off faults. The effect of stuck-at-fault resistance value and its spatial variations across the columns on the inference accuracy are also studied. An automatic Python framework is used to generate NN for software machine learning and the corresponding netlist of ReRAM neuromorphic circuit with defect distribution for SPICE simulation using Spectre. Finally, a novel scheme is proposed to recover the defect-degraded inference accuracy using machine learning.

---

*Corresponding Author: hiuyung.wong@sjsu.edu

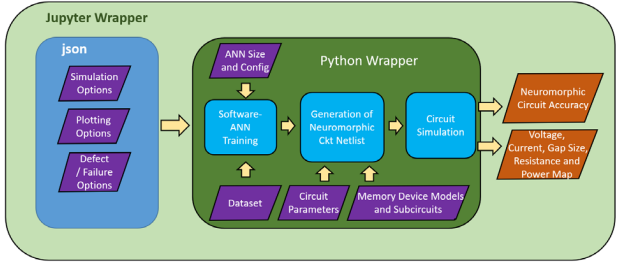

**Fig. 1.** The framework used in this study with the defect option highlighted.

## II. SIMULATION SETUP

Fig. 1 shows the framework used. It is written as Jupyter and Python wrappers. It has the following functions. Firstly, for a given problem, it constructs a neural network and performs optimization in software. It then maps the optimized software NN to netlist including mapping the weight of the NN to ReRAM conductance which is realized by changing its gap size. Users will specify the simulation conditions such as the defect size in a JSON file. It will then call a commercial software (such as Spectre) to perform circuit simulations. The jobs are submitted in parallel. After the simulation, it collects the results and calculates inference accuracy and statistics. Finally, it plots array maps of various quantities for visualization. The circuit is simulated using a generic 45nm PDK from Cadence [20]. The ReRAM is modeled using the Verilog-A model based on [21]. The model is capable of modeling the changes in gap size as a function of bias and temperature. Therefore, it captures the non-ideality during the reading of the ReRAM. This is an $HfO_x$-based ReRAM with minimum and maximum calibrated gap sizes equal to 0.2 nm (1.8mS) and 1.7 nm (4.4μS), respectively. Fig. 2 shows the ReRAM conductance versus gap size plot.

UCI Machine Learning Repository hand-written digit images are used in this study [22]. There are 1797 images. Each image has 8×8 pixels with 17 grey levels. The NN is trained with 1617 images and tested with 180 images. One of the best-trained NN has 3 hidden layers with 50, 20, and 8 nodes respectively (Fig. 3), and an inference accuracy of 96.67%. The output layer has 10 nodes and the one that gives the highest value represents the predicted number (e.g. if node 3 has the largest value, it predicts digit "3"). Note that it has 3 hidden layers. Therefore, it has 4 ReRAM arrays (array 0 to array 3) to model the synapses between the layers and perform VMM.

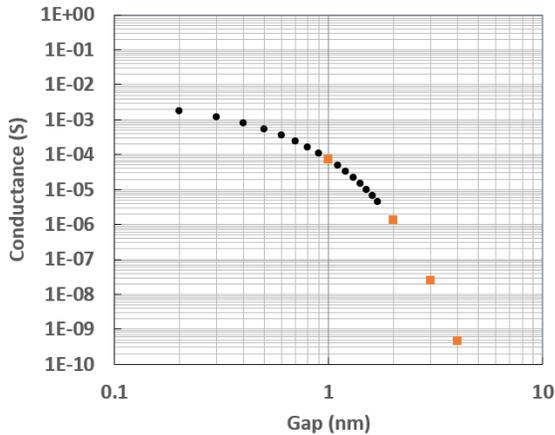

**Fig. 2** ReRAM conductance vs. gap size used in this study. Orange markers represent the gap/conductance values used in the stuck-at fault study in Fig. 7.

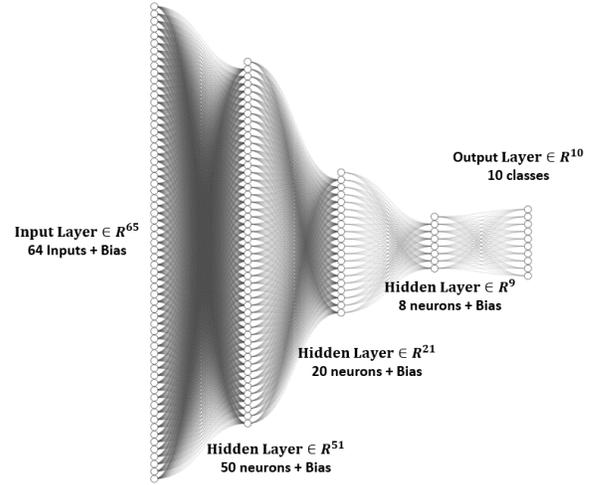

**Fig. 3** The NN studied in this paper for hand-written digit recognition.

Fig. 4 shows a simplified array structure with current comparators and rectifiers. The array and the peripheral are fully analog. Since ReRAM cannot store negative weights, two ReRAMs are needed to model negative values by comparing the relative resistance between the left and the right branches. Therefore, there are two physical columns for every logical column (e.g. C0+ and C0- form the first logical column in Fig. 4). As a result, two physical ReRAMs form one logical ReRAM in the lateral direction in the array. The output is calculated using the following equation:

$$V_{o,i} = f(I_{i+} - I_{i-}) = \sum_{i=0}^{n-1} I_{ij+} - I_{ij-}$$
$$= \sum_{i=0}^{n-1} V_{in,i}G_{ij}^+ - V_{in,i}G_{ij}^- \qquad (1)$$

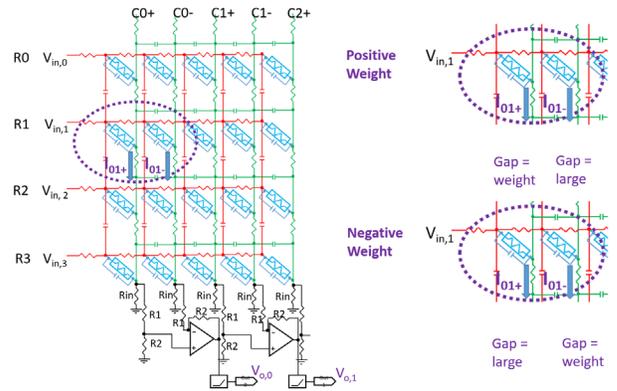

**Fig. 4** Left: A simplified netlist of the ReRAM array used to model the synapses (neuron connections). There are 4 such layers (*array 0* to *array 3*). Right: Illustration on how a logical ReRAM stores positive (top) and negative (bottom) weight using its two physical ReRAMs.

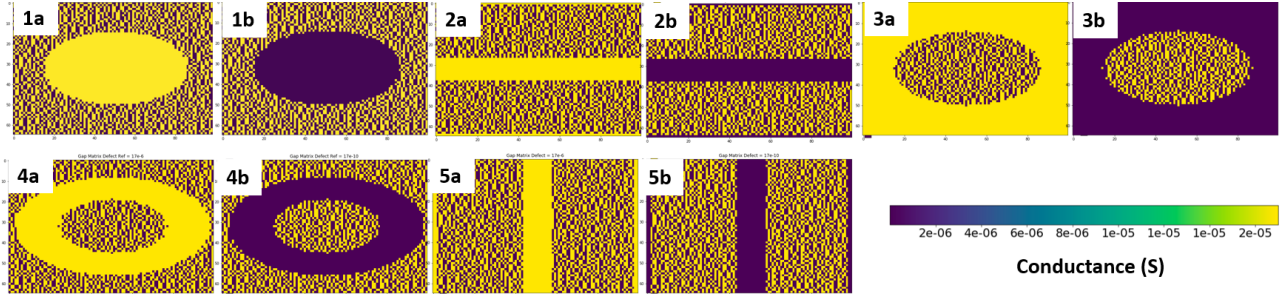

**Fig. 5** ReRAM gap size maps for various types of stuck-at fault in array 0. a's are stuck-on and b's are stuck off. 1 is circular, 2: is row, 3 is circular complement, 4 is ring, and 5 is column.

where $V_{o,i}$ is the output of column *i* logical ReRAM. It is a result of the conversion of the difference of the two physical column currents, $f(I_{i+} - I_{i-})$, where $f()$ represents the current-to-voltage converter in Fig. 4. When the weight is positive (negative), maximum gap size is assigned to the right (left) physical column and the gap corresponding to the absolute value of the weight is put on the left (right) physical column. We call the one with the maximum gap size as the *reference ReRAM*. $G_{ij}^+$ and $G_{ij}^-$ are the transconductance of the ReRAM on the left and right branches, respectively.

### III. DEFECT ASSIGNMENT

Defects with various spatial patterns, namely, circular, ring, circular-complement, row, and column defects are constructed by assigning the corresponding defective conductance in the defective region (Fig. 5). For stuck-on faults, $G_{ij}^+$ and $G_{ij}^-$ are set to 1.8mS (corresponding to a gap size of 0.2nm) in the defective region. For stuck-off faults, $G_{ij}^+$ and $G_{ij}^-$ are set to 0 mS. Various stuck-at-fault conductance values within and out of the calibration range in Fig. 2 have also been tested. Besides the aforementioned cases with $G_{ij}^+ = G_{ij}^-$, $G_{ij}^+ \neq G_{ij}^-$ in the defective region is also studied.

### IV. COMPARISON BETWEEN STUCK-ON AND STUCK-OFF FAULT

Stuck-on and stuck-off faults of various spatial patterns are added to the first ReRAM array, *array 0*, which is between the input layer and the first hidden layer of the NN. To have a fair comparison between all different types of damages, the inference accuracy as a function of the number of defective ReRAM pairs in *array 0* is plotted in Fig. 6. The total number of ReRAM pairs in *layer 0* is 3250. It is found that stuck-on and stuck-off have the same effects. This is because, in the defective region, $G_{ij}^+ = G_{ij}^-$. Based on Eq. (1), the contribution to the current will be the same regardless of the value of $G_{ij}^+$ and $G_{ij}^-$ as long as they are identical and they cancel each other ($V_{in,i}G_{ij}^+ - V_{in,i}G_{ij}^- = 0$). It is also found that most of the patterns have greater than 90% accuracy up to about 300 damaged pairs (i.e. ~10% of the total ReRAM pairs).

This type of stuck-at fault represents global damages of the array during process fabrication (e.g. etching non-uniformity in contact opening) or external mesoscopic damage during operations.

It is also possible that $G_{ij}^+ \neq G_{ij}^-$ at the defective region. That means that for the same logical ReRAM, the physical *reference ReRAM* (i.e. the one with zero conductance) and the physical ReRAM which carries the weight have different defective values. Based on the above discussion, it is expected that the inference accuracy will degrade as their contributions do not cancel each other.

Fig. 7 shows that when $G_{ij}^+ \neq G_{ij}^-$ in the defective region and when the conductance is in the operating range (between 1.8mS and 4.4μS (See also Fig. 2)), the inference accuracy degrades to be less than 60% even with a circular defect with a diameter only 10% of the size of the array, which corresponds to about 10% of the total ReRAM pairs. For conductance values outside of the normal operation range (<4.4μS), even $G_{ij}^+ \neq G_{ij}^-$, it still has a high accuracy because the terms $V_{in,i}G_{ij}^+ - V_{in,i}G_{ij}^-$ due to the defect is small and, thus, does not have a big impact.

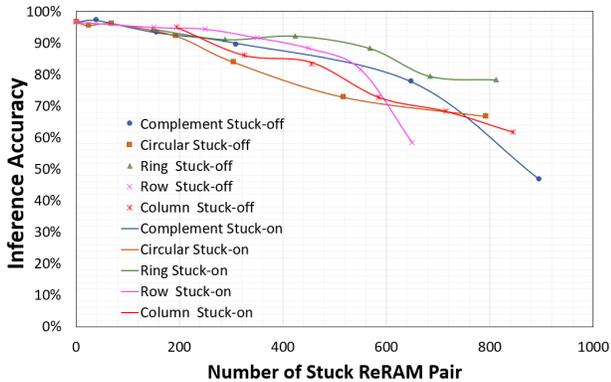

**Fig. 6** Inference accuracy as a function of the number of ReRAM pairs with stuck-off and stuck-on faults.

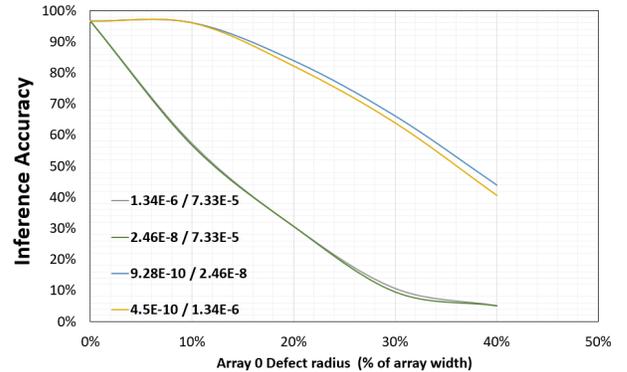

**Fig. 7** Inference accuracy of a circular defect when the defective ReRAM has different conductance in its two physical ReRAMs. The first number is the conductance of the physical ReRAM carrying the weight information and the second number is the conductance of the reference ReRAM. Unit is S.

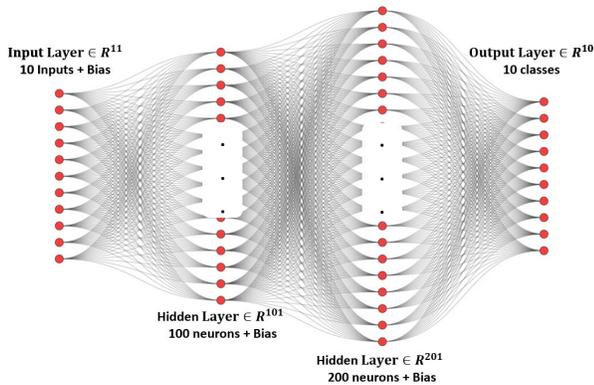

**Fig. 8** The neural network used for defect recovery. The input is the output node voltages of the neuromorphic circuit and the output is the prediction of the digit value.

| Dropout | NN Training Accuracy | NN Testing Accuracy |
|---|---|---|
| No droput | 91.43% | 83.61% |
| 10% | 86.00% | 85.00% |
| 20% | 82.79% | 84.16% |
| 30% | 83.07% | 82.50% |

**Table I:** Inference accuracy with ML correction when the ML is trained with various dropout rate.

## V. Error Correction for Array Defects

As shown in Fig. 6, even with $G_{ij}^+ = G_{ij}^-$ in the defective region, the inference accuracy drops significantly when there are more than 300 defective pairs. Since the defective pairs for a given pattern and $G_{ij}^+ = G_{ij}^-$ values influence the prediction accuracy in a predictable way, it is possible to correct the error and increase the inference accuracy by using machine learning.

Currently, the value of a given handwritten digit is determined based on the position of the output node that produces the highest voltage in the neuromorphic circuit. With defects, the highest voltage might appear in another node and cause an error. It is believed that we might be able to recover the correct value if the ten voltages are taken into consideration. Therefore, a software machine is trained using the ten voltage outputs of the neuromorphic circuit as the input features and the given digit value as the output. Fig. 8 shows the best NN constructed for a circular defect with a stuck-off fault and radius of 36.5% of the array. Among the 1797 images, 80% are used for training and 20% for testing. Before error correction, the inference accuracy is 47.78%. After being corrected by the machine, the accuracy is improved. Table I shows the improvement as a function of dropout in the training. *The highest accuracy is 85% with a 10% dropout.*

## VI. Conclusions

Using full circuit simulation, it is found that stuck-on and stuck-off faults have similar impacts on the inference accuracy of a fully analog ReRAM neuromorphic circuit if the two physical ReRAMs of a logical ReRAM have the same stuck-at-fault resistive value. If they are different, the inference accuracy is greatly degraded. We also proposed a novel scheme to recover the accuracy by training a machine using the 10 output voltages of the neuromorphic circuit as the inputs. The accuracy improved from 48% to 85% in a test example.